\documentclass[12pt]{iopart}
\usepackage{graphicx}
\begin{document}
%\bibliographystyle{prsty}
%\draft
\title[Hunting local mixmaster dynamics]{Hunting Local Mixmaster Dynamics in Spatially Inhomogeneous Cosmologies} 
%{\small}}
\author{Beverly K. Berger}
\address{Physics Division, National Science Foundation, Arlington, VA 22207 USA}

%\maketitle
%\bigskip
\begin{abstract}
Heuristic arguments and numerical simulations support the Belinskii et al (BKL) claim that the approach to the singularity in generic gravitational collapse is characterized by local Mixmaster dynamics (LMD). Here, one way to identify LMD in collapsing spatially inhomogeneous cosmologies is explored. By writing the metric of one spacetime in the standard variables of another, signatures for LMD may be found. Such signatures for the dynamics of spatially homogeneous Mixmaster models in the variables of $U(1)$-symmetric cosmologies are reviewed. Similar constructions for $U(1)$-symmetric spacetimes in terms of the dynamics of generic $T^2$-symmetric spacetime are presented. 
\end{abstract}
%\pacs{98.80.Dr, 04.20.J}
%\narrowtext

\section{Introduction}
Vincent Moncrief and I became interested in spatially inhomogeneous cosmological spacetimes as graduate students at the University of Maryland more than 30 years ago. Kuchar \cite{kuchar71} had extended Misner's minisuperspace approach to quantum cosmology \cite{misner69a} to cylindrically symmetric spacetimes containing gravitational waves (Einstein-Rosen waves \cite{einstein37}). At the same time, Gowdy had recognized that essentially interchanging the radial and time coordinates in the Einstein-Rosen solution yielded cosmological spacetimes containing gravitational waves \cite{gowdy71}. Moncrief and I had many discussions about the Gowdy models --- especially those with $T^3$ spatial topology. The classical and quantum mechanical properties of the polarized subclass of Gowdy $T^3$ models eventually became my thesis \cite{berger74}. Even then, we recognized that Gowdy spacetimes  are ideal ``theoretical laboratories" to test formalisms to develop potentially provable conjectures. Later work by a number of authors, e.g.\ \cite{isenberg90,monojlivic93},%,gowdy93,vanputten97,menamarugan97,ashtekar97,andreasson98,narita00,rendall00,torre02,chae03}, 
has demonstrated the usefulness of these models.

Both Moncrief and I returned to the Gowdy models from time to time, e.g.\ as an example of extendable spacetimes \cite{moncrief81} or quantum field theory in curved spacetime \cite{berger84}. Our paths and the Gowdy models crossed again in 1993 at (K)ITP when we used generic Gowdy $T^3$ spacetimes in the collapsing direction to investigate the approach to the singularity with numerical simulation \cite{berger93}. This project grew to involve others (for a summary see \cite{berger98c}; for a review see \cite{berger02}) and explored a variety of collapsing spatially inhomogeneous cosmologies. In this Chapter, I will discuss one aspect of my collaboration with Moncrief --- the signature of local Mixmaster dynamics (LMD) (see for example \cite{berger00}).

The singularity theorems of Penrose, Hawking, and others state that reasonable matter evolving from regular, generic initial data will develop singular or otherwise pathological behavior if the gravitational field becomes sufficiently strong. The theorems do not, however, describe the nature of these inevitable singularities. Penrose's cosmic censorship conjectures state that the singularities must be hidden from external observers by an event horizon (weak) and/or not be detectable by a timelike observer until he/she falls into it (strong). Specific spacetimes are known with many different types of singular behavior. Some yield violations of the cosmic censorship conjectures but are for various reasons viewed as non-generic. Collapsing spatially inhomogeneous cosmological spacetimes provide an arena for exploring the nature of generic singularities and for testing strong cosmic censorship. An excellent review of singularities and cosmic censorship along with references to the original papers may be found in \cite{wald84,wald97}.

Long ago, Belinskii, Khalatnikov, and Lifshitz (BKL) \cite{belinskii71b,belinskii82} had argued that the generic singularity was spacelike, local, and oscillatory. This means that, eventually, in a generic gravitational collapse, each spatial point will evolve as a separate universe. The time evolution at that point will then describe the approach to the singularity of the most general spatially homogeneous cosmology --- Bianchi Type IX (or VIII) --- the Mixmaster model \cite{misner69}. Mixmaster behavior has been a focus of study for almost 40 years (for a review see \cite{berger02}). At any instant, the state of a collapsing Mixmaster universe may be described by a single (BKL) parameter $u$ which encodes the anisotropic collapse rates of the spacetime. Each epoch of the collapse is a Kasner spacetime characterized by a fixed (in time) value of $u$. As BKL argued, this fixed value cannot be maintained as the influence of the spatial scalar curvature in a generic spacetime begins to dominate. The change in $u$ is calculated using conservation of momentum in a bounce off the dominant term in the curvature potential \cite{belinskii71b}. It is found that, for $1 \le u_n < \infty$, the value of $u$ at the $n^{th}$ epoch is given by
\begin{equation}
\label{ueq}
u_{n+1} = \left\{ {\matrix{{u_n - 1} \quad & {u_n \ge 2} \cr
{1 \over {u_n - 1}} \quad & 1 \le u_n \le 2}} \right.\ .
\end{equation}
A combination of numerical simulation \cite{berger96c} and mathematical analysis \cite{ringstrom99} %,ringstrom00} 
has demonstrated that the asymptotic dynamics approaches arbitrarily close to that described by the $u$-map (\ref{ueq}). Note that the sequence $\{ u_n \}$ is sensitive to initial conditions (due to the subtraction in the denominator). All the epochs with $u_n \to u_n - 1$ form an era. When $u_n \to (u_n - 1)^{-1}$, a new era begins. The era-to-era evolution of $u$ is described by the Gauss map $u_{N+1} = (u_N - [u_N])^{-1}$, where $[\ ]$ denotes integer part. The simple relationship of the $u$-map offers, as was known to BKL, an invariant way to characterize Mixmaster dynamics at any spatial point (local Mixmaster dynamics (LMD)). BKL referred to LMD as oscillatory behavior.

The other primary type of approach to the singularity in spatially homogeneous cosmologies is exemplified by the existence of a final Kasner (or Bianchi Type I) epoch (or $u$-value). This behavior is called asymptotic velocity term dominance (AVTD) since, once in the final Kasner epoch, the spatial scalar curvature never again plays a role in the dynamics. The BKL conjecture is  that (1) eventually all singularities will be spacelike, (2) each spatial point evolves as a separate spatially homogeneous universe, and (3) while the most general behavior is LMD, some systems will be AVTD. 

In the past few years, progress has been made in providing a rigorous underpinning to the identification of spacetimes with AVTD approaches to the singularity \cite{isenberg90}. For example, Andersson and Rendall were able to prove the existence of an open set of AVTD solutions for cosmological spacetimes on $T^3 \times R$ with no symmetries and a scalar field (to suppress LMD\footnote{It is well-known \cite{berger99a} that Mixmaster dynamics in spatially homogeneous cosmologies will be suppressed if a minimally coupled spatially homogeneous scalar field is added.}) \cite{andersson00}. Analogous constructions for spatially inhomogeneous cosmological spacetimes expected to exhibit LMD do not yet exist. (See, however, Uggla et al \cite{uggla03}). So far, the only detailed knowledge of the approach to the singularity in such models come from numerical simulations \cite{berger98c}. Numerical simulations of $T^3 \times R$ spacetimes without symmetry are in progress \cite{garfinkle03}. These use invariant quantities equivalent to a local construction of the $u$-map to search for LMD. An invariant calculation of a quantity equivalent to $u$ will yield a prediction for the next value of $u$ at that spatial point. While this approach has been tested in spacetimes of higher symmetry and preliminary results show its value in the most general case \cite{garfinkle03}, I shall focus here on another way to characterize LMD that has proven useful in these models \cite{berger00}.

Numerical simulations of Gowdy \cite{berger97b}, $T^2$-symmetric \cite{berger01}, and 
$U(1)$-symmetric spacetimes have used (in contrast to the no-symmetry case) variables adapted to the system in question. A Hamiltonian whose variation yields the relevant Einstein equations may be developed in terms of these variables and their conjugate momenta. The variables are chosen so that the Hamiltonian (density), $H$, has the form of a kinetic ``energy" plus a potential ``energy." Some of the terms in $H$ depend exponentially on the configuration variables. These, as well as non-exponential terms, may contain spatial derivatives. Neglecting both the exponential and spatial derivative terms typically yields a ``free particle" Hamiltonian. The solution to the corresponding equations of motion would describe the model's velocity term dominated (VTD) limit (if it were to exist). A heuristic argument that the model is AVTD would be that substitution of the VTD solution into the neglected terms causes them to become exponentially small as the singularity is approached. If the VTD solution is not consistent in this  way, one typically finds two or more of the exponential potentials alternately growing and decaying. This yields the (presumably) infinite sequence of bounces that characterizes LMD. This heuristic approach, called the Method of Consistent Potentials (MCP), was first introduced by Grubi\u{s}i\'{c} and Moncrief \cite{grubisic93} and later generalized \cite{berger98c}. 

Following the MCP approach, it was found that apparently different classes of potentials were important in the various non-AVTD models. Most strikingly, the Mixmaster dynamics involving the standard three minisuperspace potential terms in the spatially homogeneous models somehow become bounces off two potentials in $U(1)$-symmetric collapse and three again in generic $T^2$-symmetric collapse. To resolve this ``paradox," one can rewrite the metric of a Bianchi Type IX cosmology in the same variables as those of a spatially inhomogeneous model and thus identify the signature for Mixmaster dyanamics in these variables \cite{garfinkle01}. We note that, while we shall focus primarily on $T^3$ spatial topology, the Bianchi Type IX as $U(1)$-connection is, of course, made for $S^3$ spatial topology. Heuristic evidence is that close to the singularity, as the influence of nearby spatial points on the dynamics at any given one becomes negligible, topological differences also become unimportant. This may be demonstrated explicitly through comparison of the approach to the singularity of Gowdy models on $T^3 \times R$ \cite{berger97b} with those on $S^2 \times S^1 \times R$ \cite{garfinkle99}.

In this Chapter, I shall summarize the (published) predictions of LMD in $U(1)$-symmetric models and report on some tests of these predictions. I shall then discuss the closely related issue of expressing a $T^2$-symmetric spacetime as a $U(1)$-symmetric one. This allows similar predictions to be made from observations in $T^2$-symmetric spacetimes \cite{berger01}. We shall see that a number of interesting open questions arise in this case.

\section{Mixmaster as a $U(1)$-symmetric spacetime}
The material in this section summarizes \cite{berger00}. On $S^3 \times R$, one can compare a spatially homogeneous (diagonal) Bianchi IX metric as written in a coordinate frame:
\begin{eqnarray}
\label{metricix}
ds_{IX}^2 =& - e^{2(\alpha + \zeta + \gamma)} d\tau^2 + e^{2 \alpha} (\cos \phi \,d \theta+ \sin \theta \sin \phi \,d \psi)^2 \nonumber \\
&+e^{2 \zeta} (-\sin \phi \,d \theta + \sin \theta \cos \phi \,d \psi)^2 + e^{2 \gamma} (d \phi + \cos \theta \,d \psi)^2
\end{eqnarray}
to a spatially inhomogeneous $U(1)$-symmetric metric. $U(1)$ symmetric cosmologies on $S^3 \times R$ are described by the metric \cite{moncrief86}:
\begin{eqnarray}
\label{u1metrica}
ds_{U(1)}^2 = &e^{-2 \varphi} \{-  N^2 \, d\tau^2 + g_{ab} (dx^a +  N^a \,
d\tau)(dx^b + N^b \,d\tau) \} \nonumber \\
&+ e^{2 \varphi} (d\psi + \cos \theta \,d\phi + \beta_a
\, dx^a + \beta_0 \, d\tau)^2.
\end{eqnarray}
In (\ref{metricix}), the logarithmic scale factors (LSFs), $\alpha$, $\zeta$, and $\gamma$, are functions of $\tau$, while $\phi$, $\theta$, and $\psi$ are angles on $S^3$. In (\ref{u1metrica}), the symmetry direction is $\psi$, and the other spatial directions are $\{x^a\} =
\{\theta,\phi\}$. The metric variables are assumed to be functions of $\theta$, $\phi$,
and $\tau$. The norm of the Killing field (in the $\psi$ direction) is $e^\varphi$, $\beta_a$ are the ``twists'',
$e^{2 \Lambda}$ is the determinant of the 2-metric $ g_{ab} = e^\Lambda e_{ab}$ and
$e_{ab}$ is parametrized by $x$ and $z$ via
\begin{equation}
\label{eab}
e_{ab}={1 \over 2}\left[
{\matrix{{e^{2z}+e^{-2z}(1+x)^2}&{e^{2z}+e^{-2z}(x^2-1)}\cr
{e^{2z}+e^{-2z}(x^2-1)}&{e^{2z}+e^{-2z}(1-x)^2}\cr }} \right] \ .
\end{equation}
Note that the metric $(\ref{u1metrica})$ differs from that for $T^3$ spatial topology given
in \cite{berger98a}. It is convenient to make a canonical transformation from the twists
and their conjugate momenta $e^a$ to the twist potential $\omega$ and its conjugate
momentum $r$. It is also convenient to define the spacetime slicing by zero shift and
lapse $ N \sin \theta = \sqrt{ g} = e^\Lambda$ where $ g$ is the
determinant of the 2-metric $ g_{ab}$.
In the variables of (\ref{metricix}) , the singularity occurs at $\tau = \infty$. In any Kasner-like, epoch, as $\tau \to \infty$, two of the LSFs will be decreasing and one increasing (anisotropic collapse). In terms of the Misner variables \cite{misner69} $\Omega$, $\beta_\pm$, given by 
\begin{equation}
\label{convertbkl}
\alpha = \Omega - 2 \beta_+, \quad
\zeta = \Omega + \beta_+ + \sqrt{3} \beta_-, \quad
\gamma  = \Omega +  \beta_+ - \sqrt{3} \beta_-,
\end{equation}
the evolution of the spacetime is obtained from the Hamiltonian
\begin{equation}
\label{mixh0}
2H_{IX} = 0 = - p_\Omega^2 + p_+^2 + p_-^2 + V_{IX}(\beta_\pm, \Omega)
\end{equation}
where
\begin{equation}
\label{msspot}
V_{IX} = e^{4\alpha} + e^{4\zeta} + e^{4\gamma} - 2 e^{2(\alpha+\zeta)} - 2 e^{2(\zeta +
\gamma)} - 2 e^{2(\gamma + \alpha)}
\end{equation}
is proportional to the spatial scalar curvature and $p_\Omega$, $p_\pm$ are cannonically conjugate to $\Omega$, $\beta_\pm$. Under almost all circumstances, either $V_{IX}$ is exponentially small or one of the first three terms on the right hand side of (\ref{msspot})  dominates the dynamics. This dominant term is associated with the increasing LSF in that Kasner epoch. The increasing LSF reaches its maximum value as the corresponding momentum vanishes (and then changes sign)---i.e., a bounce off the curvature potential occurs to end the epoch. During this bounce, the more slowly decreasing LSF also has vanishing derivative at the bounce and then begins to increase. Meanwhile, the most negative LSF changes its time derivative at each bounce---continuing to decrease but more slowly. The era ends when this LSF starts to increase. 

\begin{figure}
\center
\includegraphics[scale = .5]{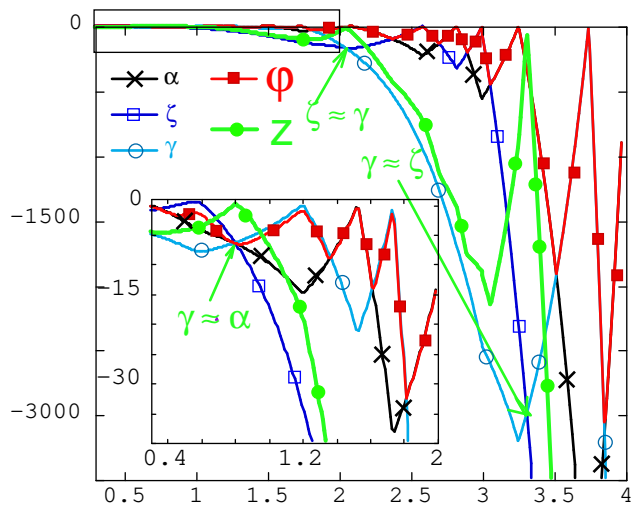}
\caption{\label{fig1}  Evolution of the Mixmaster LSFs $\alpha$, $\zeta$, and $\gamma$ toward the singularity vs $\tau$ (increasing in collapse). The inset shows the region indicated by the rectangle in the upper left hand corner. The $U(1)$ variables $\varphi$ and $z$ are also shown. Note that $\varphi$ follows the largest LSF while $z$ increases at the end of an era. The arrows show the equality of the two subdominant scale factors that marks the era's end. }
\end{figure}

Rather than the three dominant terms in $V_{IX}$, the $U(1)$-symmetric Hamiltonian whose variation yields Einstein's equations is
\begin{eqnarray}
\label{Hu1}
H_{U(1)} &=& \int_{S^3} \,{\cal H}_{U(1)} \nonumber \\
&=& \int_{S^3} {{ N e^{-\Lambda}} \over {\sin \theta}} \left[ \left( {1 \over 8}p_z^2+{1 \over 2}
e^{4z}p_x^2+{1 \over 8}p^2+{1 \over 2}e^{4\varphi }r^2-{1 \over 2}p_\Lambda
^2  \right) \right. \nonumber \\
&& + \left\{  \left( {e^\Lambda e^{ab}} \right) ,_{ab}-
\left( {e^\Lambda e^{ab}}
\right) ,_a\Lambda ,_b+e^\Lambda  \right. \left[  \left( {e^{-2z}}
\right) ,_u x,_v- \left( {e^{-2z}} \right) ,_v x,_u \right] \nonumber \\
&& \left. \left. +2e^\Lambda e^{ab}\varphi ,_a\varphi ,_b+{1 \over 2}
e^\Lambda e^{-4\varphi }e^{ab}\omega ,_a\omega ,_b \right\} \right] 
\end{eqnarray}
where ${\cal H}_{U(1)} = 0$ is the Hamiltonian constraint, the overall trigonometric factor
comes from $N/\sqrt{g}$, and $p_\varphi$, $r$, $p_\Lambda$, $p_z$, and $p_x$ are cannonically conjugate to $\varphi$, $\omega$, $\Lambda$, $z$, and $x$. The configuration variable $\omega$ is the ``twist" potential obtained from the twists $\beta_a$ in the metric (\ref{u1metric}) (see \cite{berger98a}).
Note the three ``potentials"
 \begin{equation}
\label{u1pots}
V_1 = {1 \over 2} r^2 e^{4 \varphi}, \quad
V_2 = {1 \over 2}\,e^\Lambda e^{-4 \varphi} e^{ab} \omega,_a \omega,_b \, , \quad
V_3 = {1 \over 2} p_x^2 e^{4z}.
\end{equation}
The relationship between the Bianchi IX variables $\alpha$, $\zeta$, and $\gamma$ and the $U(1)$ variables $\varphi$, $\omega$, $\Lambda$, $z$, and $x$ may be found by comparison of the corresponding metric coefficients in (\ref{metricix}) and (\ref{u1metrica}). In this summary, we shall consider only $\varphi$ and $z$.
\Fref{fig1} shows the $U(1)$ variables $\varphi$ and $z$ superposed on the LSFs obtained from a numerical simulation of a vacuum, diagonal, Bianchi IX model. Clearly, $\varphi$, given by 
\begin{equation}
\label{e2phi}
e^{2\varphi} = e^{2\alpha} \sin^2 \theta \sin^2 \phi + e^{2\zeta} \sin^2 \theta \cos^2 \phi + e^{2 \gamma} \cos^2
\theta,
\end{equation}
tracks the largest LSF. Note that $V_3$ is exponentially small unless $z$ is of order unity (rather than large and negative). Since $z$ is given by
\begin{equation}
\label{zequiv}
\fl e^{2z} ={{  4 e^{2(\alpha + \zeta +\gamma)} \sin \theta \sqrt{e^{2 \gamma} \cos^2 \theta + \sin^2 \theta (e^{2 \zeta}  \cos^2 \phi +
e^{2 \alpha} \sin^2 \phi)} } \over
{
e^{2 (\alpha+ \zeta)} + e^{2 (\alpha+ \gamma)}  + e^{2( \zeta+ \gamma)}   -e^{2 (\alpha+ \zeta)} \cos (2 \theta) + (e^{2 \alpha} -
e^{2 \zeta}) e^{2 \gamma}  \cos [2(\theta -\phi)]}}\ ,
\end{equation}
it is clear that, after factoring out the largest (i.e., expanding) LSF, $e^{2z}$ is of order unity when, at the end of an era, the now increasing smallest LSF becomes equal to the middle LSF (as seen in Figure 1). This behavior of $\varphi$ and $z$, when it appears at a given spatial point in a simulation of $U(1)$-symmetric collapse, may be thus considered a signature of LMD.

\begin{figure}
\center
\includegraphics[scale = .5]{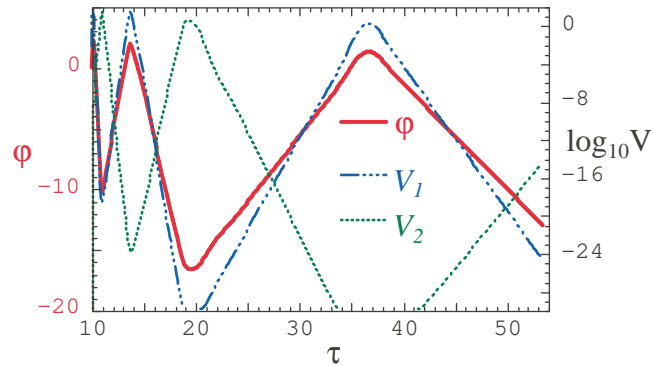}
\caption{\label{fig2}  Typical behavior of $\varphi(\tau)$ at a typical spatial point. The evolution may be characterized as bounces off the $U(1)$ potentials $V_1 = {\textstyle {1 \over 2}} r^2\,e^{4\varphi}$ and $V_2 = {\textstyle {1 \over 2}} \,e^\Lambda e^{-4\varphi}\, e^{ab}  \omega,_a \omega_b$ which are also shown. The typical behavior of $z$ (see figure 6 in \cite{berger98a}) is monotonic decrease with decreasing slope where the slope changes when $\varphi$ changes from decreasing to increasing. }
\end{figure}

\begin{figure}
\center
\includegraphics[scale = .4]{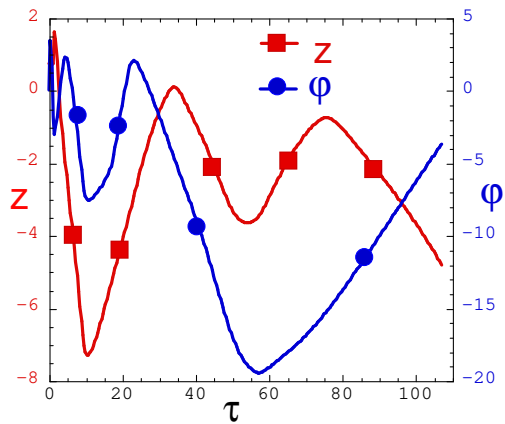}
\caption{\label{fig3}  Evidence for an era end in a $U(1)$ collapse simulation. Note that $z$ increases in association with bounces of $\varphi$ vs $\tau$ at a given spatial point. }
\end{figure}

Since such $U(1)$ simulations to date are limited in initial data {\it ansatz}, spatial resolution, and evolution time $\tau$, it is not surprising that the signature for the end of an era (increasing $z$) may not appear at all in any given simulation.  \Fref{fig2} shows the generically seen oscillatory behavior of $\varphi$ (as discussed in detail elsewhere \cite{berger98a}). The number of $\varphi$-bounces shown is typical providing further evidence that the typical evolution seen in the simulations represents the midst of an era. \Fref{fig3} shows a possible era-ending $z$-bounce from a simulation with different initial data.\footnote{The presence of two $z$-bounces in \fref{fig3} is actually consistent with the expectations of the $u$-map (\ref{ueq}). Any era with fractional part of $u$ between $.5$ and $1$ will yield a subsequent era with integer part $1$ and thus a single bounce. This makes single bounce eras quite likely.} This appearance of both $\varphi$ and $z$ bounces in the $U(1)$ collapse simulations thus provides heuristic evidence for LMD in these models.  (These $U(1)$ simulations have imposed $T^3$, rather than $S^3$, spatial topology. As was previously mentioned, one would not expect the signature for LMD to depend on spatial topology when sufficiently close to the singularity.)

\section{The vacuum Gowdy model on $T^3 \times R$ as a $U(1)$-symmetric spacetime}
The analysis in this section is based in part on some notes given to me years ago by
Moncrief.  As is apparent from chapters elsewhere in this volume, Gowdy spacetimes have proven to provide useful arenas for both numerical and mathematical studies. They may be interpreted as cosmologies with spatial topology $T^3$, $S^3$, or $S^2 \times S^1$, two spatial Killing vectors (KVs), with a natural areal time measuring the change of area in the symmetry directions. The gravitational degrees of freedom $P$ and $Q$ depend on a spatial variable $\theta$ and time $\tau$ and represent respectively the $+$ and $\times$ polarizations of gravitational waves. With 2 KVs, the Gowdy spacetimes on $T^3 \times R$ are special cases of $U(1)$-symmetric spacetimes\footnote{We shall often refer to $U(1)$-symmetric spacetimes as $U(1)$ spacetimes or models.} (with 1 spatial KV) on $T^3 \times R$. The relationship between the two was developed by Moncrief and used by us as test cases for $U(1)$ collapse simulations \cite{berger96}.

Einstein's equations for the wave amplitudes $P$ and $Q$ may be found from the Hamiltonian \cite{berger93}
\begin{equation}
\label{gowdyh}
H_G = {1 \over 2} \oint d\theta \left[\pi_P^2 + e^{-2P} \pi_Q^2 + e^{-2\tau} \left( P,_\theta^2 + e^{2P}Q,_\theta^2 \right) \right].
\end{equation}
The potentials,
\begin{equation}
\label{gowdyv1v2}
V_1 = {1 \over 2} e^{-2P} \pi_Q^2 \quad {\rm and} \quad V_2 = {1 \over 2} e^{-2\tau+2P}Q,_\theta^2,
\end{equation}
act to drive the value of $v = \left(\pi_P^2 + e^{-2P} \pi_Q^2 \right)^{1/2}$ (where $P \to v \tau$ in the asymptotically velocity term dominated (AVTD) limit as $\tau \to \infty$ \cite{berger93}) into the range $(0,1)$ consistent with the AVTD behavior of these models. (See \cite{berger97b} for details.)

Compare  the metric in \cite{berger93} for the Gowdy model:
\begin{equation}
\label{gowdymetric}
ds^2_G=-e^{\lambda /2-3\tau /2}d\tau ^2+e^{\lambda /2+\tau /2}d\theta ^2+e^{P-\tau
}(d\sigma +Q\,d\delta )^2+e^{-P-\tau }d\delta ^2
\end{equation}
to the metric \cite{berger97e} for the $U(1)$ model on $T^3 \times R$:
\begin{equation}
\label{u1metric}
ds^2_{U(1)}=-N^2e^{-2\varphi  }d\tau ^2+e^{-2\varphi  }e^\Lambda
e_{ab}\,dx^adx^b+e^{2\varphi }(dx^3+\beta _a\,dx^a)^2,
\end{equation}
with the  gauge condition $N = e^\Lambda$. Now identify the corresponding metric components.

The Killing direction $x^3$ is identified with the $\sigma$ Killing direction (identification with
$\delta$ cannot be made consistent) so that the norm of the Killing field is
\begin{equation}
\label{phidef}
2\varphi = P - \tau.
\end{equation}
The (off-diagonal) $U(1)$ ``twists'' $\beta_a$ are seen to correspond to 
\begin{equation}
\label{gowdytwists}
\beta_\delta = Q, \quad \quad \beta_\theta = 0,
\end{equation}
while a comparison of $g_{\tau \tau}$ yields $2 \Lambda - 2 \varphi  = \lambda/2
- 3 \tau/2$ to give
\begin{equation}
\label{gowdylamdef}
\Lambda = {{\lambda}\over {4 }} + {{P} \over {2}} -{{5 \tau} \over {4}}.
\end{equation}
We require that the conformal metric $e_{ab}$ have unit determinant \cite{berger97e}. Since $e_{\theta
\delta} = 0$ (there are no such cross-terms in the metric), we must have
$e_{\delta \delta} = e_{\theta \theta}^{-1}$.
The identifications
\begin{equation}
\label{gowdyethth}
e^{-2 \varphi+ \Lambda} \,e_{\theta \theta} = e^{\lambda/2 + \tau/2} \quad {\rm and } \quad 
e^{-2 \varphi  + \Lambda} \,e_{\delta \delta} = e^{-P - \tau}
\end{equation}
yield
\begin{equation}
\label{gowdyes}
e_{\theta \theta} = e^{\Lambda+2\tau}, \quad \quad e_{\delta \delta} = e^{- \Lambda-2 \tau}
\end{equation}
so that $e_{\theta \theta}\,e_{\delta \delta} = 1$ as required. This means that the Gowdy slicing is
consistent with the chosen $U(1)$ gauge condition. 

To construct $x$ and $z$, we note that the $U(1)$ conformal 2-metric is given by (\ref{eab}).
Thus, solving for $xe^{-2z}$ and $e^{-2z}$ in terms of the known components of $e_{ab}$,
we find
\begin{equation}
\label{gowdyxz}
z = - {1 \over 2} \ln \cosh (\Lambda+ 2\tau), \quad \quad x = \tanh (\Lambda+2\tau).
\end{equation}

We now wish to find the relationship between the $U(1)$ variable $\omega$ (conjugate momentum $r$)
and the Gowdy variable $Q$ (conjugate momentum $\pi_Q$).
Recall that $\beta_\delta = Q$, has conjugate momentum $e^\delta = \pi_Q$. The $U(1)$
constraint $e^a,_a = 0$ is implemented \cite{moncrief86,berger97e} by defining the twist potential $\omega$ through
\begin{equation}
\label{twistmom}
e^a = \varepsilon^{ab} \omega,_b
\end{equation}
where
\begin{equation}
\label{epsiloneq}
\varepsilon ^{ab}=\left( {\matrix{0&1\cr
{-1}&0\cr
}} \right).
\end{equation}
Therefore
\begin{equation}
\label{edelta}
e^\delta = \omega,_\theta = \pi_Q
\end{equation}
so that
\begin{equation}
\label{definew}
\omega =\int^\theta  d\theta'\,{\pi _Q}.
\end{equation}
To find the conjugate momentum $r$, we recall that the kinetic term in the $U(1)$
Hamiltonian containing $r$ is\footnote{To obtain the Hamiltonian for $T^3 \times R$ $U(1)$ models, replace $\sin \theta$ by $1$ in (\ref{Hu1}). }
\begin{equation}
\label{rterm}
{1 \over 2} r^2 \, e^{4 \varphi}
\quad
{\rm and} \quad
\omega,_\tau = r \, e^{4 \varphi}.
\end{equation}
Thus
\begin{equation}
\label{rdef}
r = e^{-4 \varphi} \int^\theta d\theta' \, \pi_Q,_\tau.
\end{equation}
But the Einstein equations for the Gowdy model yield
\begin{equation}
\label{piqdot}
\pi_Q,_\tau = e^{-2\tau} \left( e^{2P}\,Q,_\theta \right),_\theta
\end{equation}
so that
\begin{equation}
\label{requals}
r = Q,_\theta.
\end{equation}
The relevant $U(1)$ kinetic term is then
\begin{equation}
\label{u1kin}
{1 \over 2} r^2 \, e^{4 \varphi} = {1 \over 2} (Q,_\theta)^2 \, e^{2P-2\tau}
\end{equation}
which is the Gowdy potential $V_2$ in (\ref{gowdyv1v2}). The $U(1)$ curvature potential term is
\begin{equation}
\label{u1curvterm}
{1 \over 2} e^\Lambda \,e^{-4\varphi}\, e^{\theta \theta} (\omega,_\theta)^2. 
\end{equation}
But $e^{\theta \theta} = e^{-\Lambda-2 \tau}$ so that (\ref{u1curvterm}) becomes
\begin{equation}
\label{gowdyv1}
{1 \over 2} \pi_Q^2\,e^{-2P}
\end{equation}
which is precisely the Gowdy potential $V_1$ in (\ref{gowdyv1v2}).  Note that the canonical transformation $(\beta_a, e^a) \to (r,\omega)$ has interchanged the roles of $V_1$ and $V_2$.

The remaining momenta are
\begin{equation}
\label{gowdypz}
p_z = 4 z,_\tau = - 2 \tanh (\Lambda+2\tau) \,( \Lambda,_\tau,+2),
\end{equation}
\begin{equation}
\label{gowdypx}
p_x = e^{-4z}\,x,_\tau = (\Lambda,_\tau+2),
\end{equation}
\begin{equation}
\label{gowdypl}
p_\Lambda = - \Lambda,_\tau.
\end{equation}
The terms in the $U(1)$ Hamiltonian containing these momenta are
\begin{equation}
\label{gowdypzterm}
{1 \over 8} p_z^2 = {1 \over 2} \,\tanh^2 (\Lambda + 2\tau) \, (\Lambda,_\tau+2)^2
\end{equation}
and
\begin{equation}
\label{gowdypxterm}
{1 \over 2} p_x^2 \, e^{4z} = {1 \over 2}\, {1 \over {\cosh^2( \Lambda+2\tau)}}\,
(\Lambda,_\tau+2)^2
\end{equation}
so that
\begin{equation}
\label{gowdypxpzcombined}
{1 \over 8} p_z^2 + {1 \over 2} p_x^2 \, e^{4z} = {1 \over 2} (\Lambda,_\tau+2)^2 .
\end{equation}

Note that if $\Lambda \to \pm \infty$ linearly in $\tau$ as $\tau \to \infty$ (as expected
asymptotically), $p_z \to -2 (\Lambda,_\tau + 2)$, a constant, while $e^{4z} \to 0$. These
behaviors are consistent with the MCP analysis of Gowdy spacetimes.

\section{Generic $T^2$ symmetric spacetimes as $U(1)$-symmetic models}
The Gowdy spacetimes are not the most general $T^3 \times R$ spacetimes with two spatial KVs. As recognized by Gowdy \cite{gowdy71}, the most general $T^2$-symmetric spacetimes have additional off-diagonal "twist" metric components. In the vacuum case, the information in the twist terms may be reformulated as twist constants (in time and space). Without loss of generality, one twist constant may be set equal to zero. The $T^2$-symmetric spacetimes have been studied analytically \cite{berger97} and numerically \cite{berger01}. Generic models appear heuristically to exhibit LMD with bounces off generalizations of the Gowdy potentials $V_1$ and $V_2$ and off a new twist potential $V_3$. Einstein's equations may be found from the variation of the Hamiltonian density
\begin{eqnarray}
\label{hgal}
{\cal H}_{T^2} &=& {1 \over {4 \pi_\lambda}} \left[\pi_P^2 + e^{-2P} \pi_Q^2 + e^{-2\tau} \left( P,_\theta^2 + e^{2P} Q,_\theta^2 \right) \right] \nonumber \\
&&+ \sigma \, \kappa^2 \, \pi_\lambda
e^{(\lambda + 2P + 3\tau ) / 2} .
\end{eqnarray}
In addition, there is a nontrivial momentum constraint:
\begin{equation}
\label{t2mom}
\pi_P \,  P,_\theta + \pi_Q \,  Q,_\theta +
\pi_\lambda \, \lambda,_\theta  =  0.
\end{equation}
Details may be found in \cite{berger01}.  $T^2$-symmetric spacetimes are a special case of $U(1)$-symmetric spacetimes.

The metric for the $T^2$-symmetric spacetimes is given by \cite{berger01}
\begin{eqnarray}
\label{galmetric}
ds^2_{T^2}&=&-e^{(\lambda -3\tau )/2}d\tau ^2+e^{(\lambda +\mu +\tau )/2}d\theta
^2+\sigma e^{P-\tau }[dx+Qdy \nonumber \\
 & +&(G_1+QG_2)d\theta -(M_1+QM_2)e^{-\tau }d\tau ]^2 \nonumber \\
  &+&\sigma e^{-P-\tau }(dy+G_2d\theta -M_2e^{-\tau }d\tau )^2
\end{eqnarray}
where the Gowdy metric  (\ref{gowdymetric}) is recovered if the shifts, $M_a$, and twists, $G_a$, vanish
and $\mu$ and $\sigma$ are set $=1$. The $G_a$'s and $M_a$'s are related to the twist
constant $\kappa$ through
\begin{equation}
\label{twist1}
e^{-\tau}M_1,_\theta + G_1,_\tau = \kappa Q e^{(\lambda + 2P + 3\tau)/2} e^{\mu/4},
\end{equation}
\begin{equation}
\label{twist2}
e^{-\tau}M_2,_\theta + G_2,_\tau = - \kappa e^{(\lambda + 2P + 3\tau)/2} e^{\mu/4}.
\end{equation}
This metric may be simplified (without loss of generality) by following a procedure introduced by Weaver \cite{weaver99} to eliminate the shifts. It is also convenient to revert to $\sigma$ and $\delta$ as coordinates. To
avoid confusion with the coordinate of the same name (and because this is done in our computer simulations), set $\sigma = 1$
in (\ref{galmetric}). Let
\begin{equation}
\label{newx}
\sigma = x - \int^\tau d\tau' \,\, M_1\,e^{-\tau'},
\quad
\delta = y - \int^\tau d\tau' \, \, M_2\,e^{-\tau'},
\end{equation}
and define
\begin{equation}
\label{thetadef}
\Theta = \kappa e^{(\lambda + 2P + 3 \tau)/2} e^{\mu/4} .
\end{equation}
If these transformations are substituted into (\ref{galmetric}), we obtain the same
results as if we demand that the shifts vanish---i.e.~the $M_a = 0$ in (\ref{twist1}) and
(\ref{twist2}). The latter condition implies
\begin{equation}
\label{twistreplace}
G_1 = \int^\tau d\tau' \, \,(Q \Theta), \quad \quad G_2 = - \int^\tau d\tau'\,\, \Theta
\end{equation}
so that (\ref{galmetric}) becomes
\begin{eqnarray}
\label{galmetricnew}
ds^2_{T^2}&=&-e^{(\lambda -3\tau )/2}d\tau ^2+e^{(\lambda +\mu +\tau )/2}d\theta
^2 \nonumber \\
&+&e^{P-\tau }\left\{ d\sigma+Qd\delta +\left[ \int^\tau d\tau'\, (Q \Theta) - Q \int^\tau d\tau'\, \Theta\right]\,d\theta\right\}^2
\nonumber
\\
   &+& e^{-P-\tau }[d\delta-(\int^\tau  d\tau'\,\Theta)\,d\theta]^2.
\end{eqnarray}
From (\ref{galmetricnew}), we can identify the $U(1)$ twists as
\begin{equation}
\label{galbetadelta}
\beta_\delta = Q,
\end{equation}
\begin{equation}
\label{galbetatheta}
\beta_\theta = \int^\tau d\tau'\, (Q \Theta) - Q \int^\tau d\tau'\, \Theta.
\end{equation}
Note that as $\tau \to \infty$, $\beta_\theta \to 0$ since $Q$ becomes constant in $\tau$.
For future reference, we note that
\begin{equation}
\label{twistdots}
\beta_\delta,_\tau = Q,_\tau, \quad \quad \beta_\theta,_\tau = - Q,_\tau \,\int^\tau d\tau'\,
\Theta.
\end{equation}
We shall see below that these assignments are consistent with the relation between the $U(1)$ twists
and $\omega$ given most simply by \cite{berger97e} 
\begin{equation}
\label{twistrule}
\beta_a,_\tau = - N e^{-4 \varphi} e_{ab} \varepsilon^{bc} \omega,_c.
\end{equation}
As in the Gowdy case, we identify (see (\ref{edelta})) $\omega,_\theta = \pi_Q$.

Now consider the 2-metric $g_{ab}$ which we define here to include all factors in
(\ref{galmetricnew}):
\begin{equation}
\label{gthth}
g_{\theta \theta} = e^{(\lambda + \mu + \tau)/2} + \left( \int^\tau \Theta
\right)^2\,e^{-P-\tau},
\end{equation}
\begin{equation}
\label{gdeldel}
g_{\delta \delta} = e^{-P-\tau},
\end{equation}
\begin{equation}
\label{gdelth}
g_{\delta \theta} = - e^{-P-\tau} \, \int^\tau \Theta.
\end{equation}
Note that
\begin{equation}
\label{det2g}
g_{\theta \theta}\,g_{\delta \delta}-g_{\delta \theta}^2 = e^{(\lambda + \mu + \tau)/2}
\equiv  e^{2 \Lambda - 4 \varphi }
\end{equation}
so that (\ref{det2g}) defines $\Lambda$.

Since the norm of the Killing field in the $\sigma$ direction is $g_{\sigma \sigma} =
e^{P-\tau}$, we identify
\begin{equation}
\label{galphidef}
2 \varphi = P - \tau
\end{equation}
as in the Gowdy case. Combining (\ref{det2g}) and (\ref{galphidef}) yields
\begin{equation}
\label{gallamdef}
\Lambda = {{\lambda} \over 4} + {P \over 2} - {{5 \tau} \over 4} + {\mu \over 4}.
\end{equation}
Note that the identification of $\Lambda$ is different from that in the Gowdy case.

We also note that this implies that the $U(1)$ gauge condition used previously and in \cite{berger98a} must be generalized. A
comparison of (\ref{u1metric}) and (\ref{galmetricnew}) yields
\begin{equation}
\label{ndef}
N = e^{(\lambda + 2P -5 \tau)/4} = e^{-\mu/4}\,e^\Lambda
\end{equation}
so that $N \ne e^\Lambda$ which violates the previously used $U(1)$ gauge condition. In the following,
(\ref{ndef}) shall be used as the gauge condition.

The generalized $T^2$ symmetric models satisfy \cite{berger01}
$\pi_\lambda = {\textstyle {1 \over 2}} e^{\mu/4}
$
so that
$N = {{e^\Lambda} / ({2 \pi_\lambda})}$.
This choice of lapse will provide the needed $\pi_\lambda$'s in the denominator of the Hamiltonian  density (\ref{hgal}) and
strongly suggests that a similar gauge condition would be needed to describe any
generalization of Gowdy spacetimes (e.g. magnetic Gowdy \cite{weaver98}) as $U(1)$ models.

Now identify the components of the conformal metric $e_{ab}$ using 
$g_{ab} = e^{\Lambda - 2\varphi } e_{ab}
$
to find
\begin{equation}
\label{galethth}
e_{\theta \theta} = e^{(\lambda + 2P + 3 \tau + \mu)/4} +e^{-(\lambda + 2P + 3 \tau +
\mu)/4} \left(\int^\tau d\tau'\, \Theta \right)^2,
\end{equation}
\begin{equation}
\label{galedeldel}
e_{\delta \delta} = e^{-(\lambda + 2P + 3 \tau +\mu)/4},
\end{equation}
\begin{equation}
\label{galedelth}
e_{\delta \theta} = -e^{-(\lambda + 2P + 3 \tau +
\mu)/4}\, \int^\tau d\tau'\, \Theta.
\end{equation}
We shall also need 
$e^{\theta \theta} = e^{-(\lambda + 2P + 3 \tau +\mu)/4}$.
Using (\ref{eab}) and (\ref{gallamdef}) with (\ref{galethth})-(\ref{galedelth}) gives
\begin{equation}
\label{galz}
e^{-2z} = e^{\Lambda+2\tau} + e^{-\Lambda-2\tau}\, \left(1 + \int^\tau  d\tau'\,\Theta \right)^2
\end{equation}
and
\begin{equation}
\label{galx}
x = {{e^{\Lambda+2\tau} - e^{-\Lambda-2\tau} \left[ 1 - \left( \int^\tau \Theta \right)^2 \right]} \over
{e^{\Lambda+2\tau} + e^{-\Lambda-2\tau}\left(1 + \int^\tau \Theta \right)^2}}.
\end{equation}
Now consider the $Q$ degree of freedom. From (\ref{twistrule}), we require
\begin{equation}
\label{betathrule}
\beta_\theta,_\tau = -  N e^{-4\varphi}e_{\theta \delta} (-\omega,_\theta)
 \quad {\rm and} \quad
\beta_\delta,_\tau =  -  N e^{-4\varphi}e_{\delta \delta} \omega,_\theta. 
\end{equation}
Our previous identifications for the $U(1)$ twists in (\ref{galbetatheta}) and
(\ref{galbetadelta}) and our identifications of the other variables yield the required
relations
\begin{equation}
\label{betadots1}
Q,_\tau = e^{-2P} {{\pi_Q} \over {2 \pi_\lambda}}
\quad {\rm
and} \quad
-Q,_\tau \int^\tau  d\tau'\,\Theta = -e^{-2P} {{\pi_Q} \over {2 \pi_\lambda}}\int^\tau  d\tau'\,\Theta .
\end{equation}

The term containing $r$ in the Hamiltonian becomes (with our new lapse condition)
\begin{equation}
\label{galrterm}
{N \over {e^\Lambda}} {1 \over 2} r^2 e^{4\varphi} = {{r^2} \over {4 \pi_\lambda}}e^{4
\varphi}
\end{equation}
so that the variation yields
\begin{equation}
\label{galrdef}
r = 2 \pi_\lambda e^{-4\varphi} \int^\theta d\tau'\, \pi_Q,_\tau.
\end{equation}
But, in these models,
\begin{equation}
\label{galpiqdot}
\pi_Q,_\tau = e^{-2\tau}\, {\partial \over {\partial \theta}} \left( {{e^{2P} Q,_\theta}  \over {2 \pi_\lambda}} \right)
\end{equation}
so that $r = Q,_\theta$ as in the Gowdy case and 
\begin{equation}
\label{finalrterm}
{{r^2} \over {4 \pi_\lambda}} e^{4\varphi} = {{e^{2(P-\tau)}} \over {4 \pi_\lambda}}
(Q,_\theta)^2.
\end{equation}
The other Gowdy-like potential term is
\begin{equation}
\label{galv2}
(N e^\Lambda) e^{-\Lambda} e^{\theta \theta} e^{-4\varphi} (\omega,_\theta)^2.
\end{equation}
Following the steps as with the Gowdy model  (since $e^{\theta \theta}$ is the same as in that
case) yields for this term
\begin{equation}
\label{galv2final}
{{\pi_Q^2} \over {4 \pi_\lambda}} e^{-2P}.
\end{equation}
Since the two terms in the Hamiltonian containing $Q$ arise here essentially as in the Gowdy case, we
must look elsewhere for the twist contribution.

Let us construct
\begin{equation}
\label{fdef}
F = {N \over {e^\Lambda}} \left( {1 \over 8} p_z^2 + {1 \over 2} p_x^2 e^{4z} \right)
\end{equation}
from the $U(1)$ Hamiltonian. Note that our gauge condition means that $N /{e^\Lambda} =
e^{-\mu/4} = 1/(2 \pi_\lambda)$. Since $z$ and $x$ are known, the variation of $F$ yields
expressions for $p_z$ and $p_x$ in terms of $z,_\tau$ and $x,_\tau$. We find
\begin{eqnarray}
\label{galpx}
p_x =& -{1 \over 2} e^{-2 \Lambda - 4\tau + \mu/4} \left[ \left( -1 + e^{2 \Lambda+4\tau} - 2 \int^\tau d\tau'\,
\Theta - (\int^\tau d\tau'\, \Theta )^2 \right) \, \Theta \right. \nonumber \\
&  \left.- 2 e^{2 \Lambda+4\tau} \left(1 + \int^\tau d\tau'\,
\Theta \right) \, (\Lambda,_\tau+2) \right]
\end{eqnarray}
\begin{equation}
\label{galpz}
\fl p_z = - {{2 e^{\mu/4} \left[ 2 \left( 1 + \int^\tau d\tau'\, \Theta \right) \, \Theta + 
\left( -1 + e^{2 \Lambda} - 2 \int^\tau d\tau'\,
\Theta - (\int^\tau d\tau'\,\Theta )^2 \right) \, \Lambda,_\tau \right]} \over 
{\left( -1 + e^{2 \Lambda} + 2 \int^\tau d\tau'\,
\Theta + (\int^\tau d\tau'\,  \Theta )^2 \right)} } \quad .
\end{equation}
Remarkably, while neither term on the right hand side of (\ref{fdef}) looks simple, we find that
\begin{equation}
\label{feval}
F = {1 \over 2} e^{-2 \Lambda -4 \tau + \mu/4} \Theta^2 + {1 \over 2} e^{\mu/4} (\Lambda,_\tau+2)^2.
\end{equation}
The second term on the right hand side of (\ref{feval}) will reproduce the appropriate generalization
of the comparable Gowdy term (\ref{gowdypzterm}). On the other hand, substitution of
(\ref{thetadef}), (\ref{det2g}), and the definition of $\mu$ in terms of $\pi_\lambda$ gives
\begin{equation}
\label{twistpot}
{1 \over 2} e^{-2 \Lambda+4\tau + \mu/4} \Theta^2 = \pi_\lambda \kappa^2 e^{(\lambda + 2P + 3\tau)/2}
\end{equation}
which is precisely the twist potential. This means that twist bounces (i.e.\ bounces off the twist potential) are associated with $z$-bounces in these models.

\section{Discussion}
\begin{figure}
\center
\includegraphics[scale = .4]{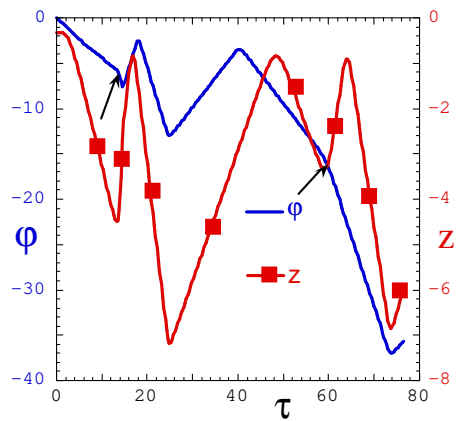}
\caption{\label{fig4}  Generic $T^2$-symmetric collapse simulation in terms of $U(1)$ variables $\varphi$ and $z$ at a representative spatial point. The arrows mark the twist bounces (see \cite{berger01} for details.) \Eref{galphidef} defines $\varphi$ in terms of the $T^2$-symmetric variable $P$ and essentially follows its behavior. Note that $z$ starts to increase at a twist bounce as well as at the end of an era.}
\end{figure}

In Section 2, we reviewed the analogies between Bianchi IX and $U(1)$-symmetric models to argue that ``$z$-bounces" should occur in $U(1)$ models at any spatial point that experiences the end of a BKL era. Numerical evidence was presented in \fref{fig2}. Here we look for the analog of twist bounces \cite{berger01} in $U(1)$-symmetric models. \Fref{fig4} shows $\varphi$ and $z$ (as functions of $P$, $Q$, $\lambda$, and $\tau$) from a $T^2$ simulation with the twist bounce marked. Thus the signature for a twist bounce in a $U(1)$ model should be a change in the slope of $\varphi$ accompanied by a change in $z$ from decreasing to increasing. 

To explore $U(1)$ simulations for these features, we must first consider the gauge condition.
To understand the difference between the our original gauge condition, $N = e^\Lambda$
\cite{berger97e} and our revised gauge condition $ N = e^\Lambda/(2
\pi_\lambda)$ (\ref{ndef}), we digress to the homogeneous cosmology case. Recall that for
anisotropic LSFs $\alpha$, $\zeta$, $\gamma$, we define $3\Omega = \alpha + \zeta +\gamma$. Einstein's
equation for $\Omega$ is
\begin{equation}
\label{wdot}
\dot \Omega = - {{\cal N} \over {\sqrt {{}^3\/g}}}\,p_\Omega.
\end{equation}
Here we distinguish the 3-D lapse ${\cal N}$ from the 2-D lapse $N$  used in
discussion of $U(1)$ models. In the homogeneous models, the BKL time $\tau$ is defined by
the gauge condition ${\cal N}= \sqrt {{}^3g}$. This is equivalent to the original gauge
condition since, from the $U(1)$ metric (\ref{u1metric}), 
${\cal N} = e^{-\varphi } N
$
while
$
\sqrt {{}^3\/g} = e^{-\varphi } \,e^\Lambda
$
so that the 3-D gauge condition is maintained.

On the other hand, if we wish to use a geometrical (e.g. volume) time coordinate, then
(e.g.) we may choose $\dot \Omega = -1$ which is equivalent to the gauge condition
\begin{equation}
\label{arealgauge}
\tilde {\cal N} = e^{-\varphi }/p_\Omega. 
\end{equation}
If $p_\Omega$ is a constant, the two gauge conditions are equivalent. 

The Gowdy areal time choice is then comparable to the volumetric gauge condition
(\ref{arealgauge}). (Note that in \cite{berger74} it is shown that writing (polarized)
Gowdy models as inhomogeneous generalizations of Bianchi I cosmologies in Misner's minisuperspace
variables requires a rotation such that $-p_\Omega^2 + p_+^2 \to - p_\tau p_\lambda $ so
that the Gowdy spacetime equivalent of (\ref{wdot}) will be $\dot \tau = ({\cal N}/\sqrt{{}^3 g}\,\,
\pi_\lambda$.) Since $\pi_\lambda$ is constant in the Gowdy spacetime, the areal time choice is
compatible with the BKL gauge condition. Once we consider generalized $T^2$-spacetimes,
the BKL gauge condition is no longer compatible and we must use the areal gauge condition
(\ref{ndef}) which is the appropriate analog of (\ref{arealgauge}).

\begin{figure}
\center
\includegraphics[scale = .4]{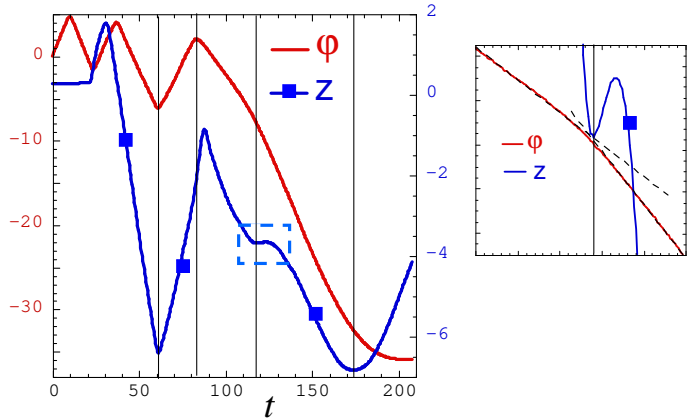}
\caption{\label{fig5}  A possible twist bounce in a $U(1)$ collapse simulation using the modified gauge condition $N=e^\Lambda / p_\Lambda$. The $U(1)$ variables $\varphi$ and $z$ vs $\tau$ are shown at a single spatial point. The inset displays the feature marked by the rectangle and shows a change in the slope of $\varphi$ along with a change in $z$ from decreasing to increasing, the signature for LMD in $T^2$-symmetric collapse. The vertical lines are used to indicate the alignment  of features in $\varphi$ and $z$.}
\end{figure}

To better search for twist bounces in $U(1)$ models, we replace the gauge condition $N = e^{\Lambda}$ with $N = e^\Lambda / p_\Lambda$ as an analog of (\ref{arealgauge}). \Fref{fig5} shows a possible twist bounce in such a simulation although one cannot regard the evidence to be compelling. The required signature is present although at a barely detectable scale.

Nonetheless, we may argue that this approach of defining signatures of LMD in simpler models has proven (and will prove) useful. Future work consists of more careful analysis of improved $U(1)$ simulations as well as the development of a direct connection between non-diagonal Bianchi IX \cite{ryan75} and $T^2$-symmetric models. Discussion in \cite{berger01} indicates that the LMD seen in $T^2$-symmetric models can be understood as requiring a bounce off a rotational potential wall. Such a wall requires generalization of the metric in (\ref{metricix}) and \cite{berger00} to include non-diagonal metric components for the metric expressed in spatial 1-forms.

The need for the new gauge condition and the need for a more general Mixmaster model may imply that the $U(1)$ metric used in \cite{berger98a} should be generalized. The search for LMD signatures can provide heuristic evidence for or against the generality of these simulations.

\section*{Acknowledgments}
Some of the results reported here were based on earlier work supported in part by NSF Grant PHY-9800103  to Oakland University and PHY-9407194  to the University of California at Santa Barbara. I would like to thank the Albert Einstein Institute (Golm, Germany), the Institute for Geophysics and Planetary Physics at Lawrence Livermore National Laboratory, and the Kavli Institute for Theoretical Physics for hospitality. I would also like to thank David Garfinkle, Jim Isenberg, and Marsha Weaver for useful discussions. Finally, I would like to express very special thanks to Vincent Moncrief for his insights and encouragement in this project and many others.

%\end{references}

\end{document}